\documentstyle[multicol,aps,psfig]{revtex}
\renewcommand{\narrowtext}{\begin{multicols}{2}
\global\columnwidth20.5pc\noindent}
\renewcommand{\widetext}{\end{multicols}
\global\columnwidth42.5pc}
\begin{document}
\draft
\preprint{September 2000}
\title{
An Electron Spin Resonance Selection Rule for Spin-Gapped Systems
}
\author{
T\^oru {\sc Sakai}$^{1,2}$, Olivier {\sc C\'epas}$^2$ and Timothy
{\sc Ziman}$^{2,3}$
}
\address
{
$^1$Faculty of Science, Himeji Institute of Technology,
3-2-1 Kouto, Kamigori-cho, Ako-gun, Hyogo 678-1297, Japan \\
$^2$Institut Laue Langevin, B.P. 156 38042 Grenoble Cedex 9, France \\
$^3$Laboratoire de Physique et Mod\'elisation des Milieux Condens\'es,
CNRS-UMR5
493 , rue des Martyrs, Grenoble, France
}
\date{\today}
\maketitle
\begin{abstract}
The direct electron spin resonance (ESR) absorption between a singlet
ground state and the triplet excited states of spin gap systems is
investigated.  Such an absorption, which is forbidden by the
conservation of the total spin quantum number in isotropic
Hamiltonians, is allowed by the Dzyaloshinskii-Moriya interaction.  We
show a selection rule in the presence of this interaction, using the
exact numerical diagonalization of the finite cluster of the
quasi-one-dimensional bond-alternating spin system. The selection rule
is also modified into a suitable form in order to interpret recent
experimental
results on CuGeO$_3$ and NaV$_2$O$_5$.
\end{abstract}
\pacs{KEYWORDS: spin gap, ESR, Dzyaloshinskii-Moriya interaction}
\narrowtext

The  spin gap generated by strong quantum fluctuations
in  low-dimensional systems, such as the bond
alternating chain, the spin 1-chain or spin ladders, continues
to attract
much  interest, both experimental and theoretical. Electron spin
resonance (ESR)
experiments at very low temperatures provide very high-energy
resolution measurements of the spin gap. Tuning an external magnetic
field, the field-dependent energy of the excited triplet state is
adjusted to the frequency of the electromagnetic propagating wave. A
direct transition from the singlet ground state to the gapped triplet
state may be observed. Such a direct transition would be, however,
forbidden by the conservation law of the total spin quantum number for
a perfectly isotropic but gapped spin system.

In recent ESR measurements for the spin-Peierls system
CuGeO$_3$\cite{nojiri} and the bond-alternating chain system
NaV$_2$O$_5$\cite{nojiri1}, direct transitions from the ground state
to the first excited triplet state have been detected.  Since the
observed intensity of the transition for CuGeO$_3$ was almost
independent of the external magnetic field, Nojiri et al.\cite{nojiri}
concluded that the explanation may be provided by the
Dzyaloshinskii-Moriya (DM)
 interaction, which had
been suggested on the basis of electron paramagnetic resonance
measurements\cite{yamada}. In contrast, another
mechanism based on the effective staggered field due to alternating
g-tensors\cite{sakai}, which had successfully explained the ESR
absorption of the Haldane gap for the $S=1$ antiferromagnetic chain
Ni(C$_2$H$_8$N$_2$)$_2$NO$_2$(ClO$_4$),
abbreviated NENP,
should lead to a strong field dependence of the intensity.  In
addition, Kokado and Suzuki's phenomenological Hamiltonian including
the DM interaction reproduced some of the results from the
angle-dependent ESR measurement of CuGeO$_3$\cite{kokado}. These works
motivated the derivation of a selection rule for direct ESR absorption
in
the presence of the DM interaction.  In the present letter, we present
such a general rule which emphasizes the role of the relative
orientations of the external field, the
{\it polarized} propagating magnetic
field and the vector of the DM interaction.
We also present a more suitable form for the experimental results, for
which the electromagnetic wave is
{\it not polarized}.

When the magnetic field of the wave ${\bf h}(t)$ is polarized
along an axis $\alpha$  $(\alpha=x,y,z)$, the intensity of ESR for
quantum
spin systems at zero temperature, assumed due to the magnetic
dipole transitions,  is determined   by the matrix element
\begin{eqnarray}
I^{\alpha}= |\langle g|\sum_j S_j^{\alpha}|e \rangle |^2,
\label{intensity}
\end{eqnarray}
where $S_j^{\alpha}$ is the $\alpha$-component of the local spin
operator at the $j$-th site, $|g\rangle$ and $|e\rangle$
denote the ground state and the excited state with the energy
difference $E_e-E_g$ corresponding to the frequency
$\omega$ of the electromagnetic wave.  For an isotropic spin
system, the dynamics  conserves the total spin quantum number
$S_{\rm total}$.  Therefore, every component of the intensity
(\ref{intensity}) vanishes.  In the presence of the DM interaction,
however, $S_{\rm total}$ is no longer a good quantum number and some
intensities may become finite.  To investigate the effect of the DM
interaction on the intensity, we first consider the problem of two
coupled $S=1/2$ spins with the
Hamiltonian
\begin{eqnarray}
{\cal H}_{(2)}= J {\bf S}_1\cdot {\bf S}_2+
{\bf D}\cdot ({\bf S}_1\times {\bf S}_2)
-{\bf H}\cdot ({\bf S}_1+{\bf S}_2),
\label{2spin}
\end{eqnarray}
where ${\bf D}$
is the vector of the DM interaction and ${\bf H}$ is
the external magnetic field.
Solving the  problem of two spins,
we calculate the intensities of the ESR transitions from
the ground state to the excited states.
%
%
For ${\bf H}=0$ and ${\bf D}\parallel {\bf z}$, they are given by:
\begin{eqnarray}
I^z=\mid \langle g, S^z=0 \mid \sum_j S_j^z \mid e, S^z=0 \rangle
\mid^2= 0
\label{r2spinz}
\end{eqnarray}
\begin{eqnarray}
I^x=\mid \langle g, S^z=0 \mid \sum_j S_j^x \mid e, S^z=\pm 1 \rangle
\mid^2= \frac{1}{8} \left(\frac{D^z}{J}\right)^2.
\label{r2spinx}
\end{eqnarray}
For nonzero magnetic field, this leads to the following selection rule.
Nonzero intensity appears only when the ${\bf h}(t)$ has
some component satisfying one of the following two conditions:
\begin{flushleft}
(i) ${\bf h}(t)
\perp {\bf H} \parallel {\bf D}$: the intensity is independent
of $H$.\\
%
%
(ii) ${\bf h}(t)
\parallel {\bf H} \perp {\bf D}$: the intensities depend on
$H\equiv |{\bf H}|$ and the
intensity of the transition to the nondegenerate state
($S^z_{\rm total}=0$ at $H=0$), which vanishes at $H=0$ increases as
$H^2$ for small $H$.
For the degenerate states ($S^z_{\rm total}=\pm1$ at $H=0$), the
intensities decrease.
\end{flushleft}
\noindent
On the other hand, the intensity would
still vanish if ${\bf h}(t) \parallel {\bf H} \parallel {\bf D}$
because of the rotational symmetry around these vectors.  Since these
rules depend only on the relative configuration of the vectors ${\bf
h}(t)$, ${\bf H}$ and ${\bf D}$, and they are independent of the
lattice structure, they are expected to be valid even in general bulk
systems with a spin gap.
%
%
(The $H$-independent intensity of rule (i) is generally justified
by the conservation of $S^z_{\rm total}$.)
We will justify them with the following
numerical analysis on  finite clusters of the quasi-one-dimensional
bond-alternating spin system. In particular cases, special symmetries
may lead to vanishing intensities at certain frequencies, in which case
the
gap observed in ESR may not coincide with that of neutron scattering.

In order to examine the above selection rules in many-spin systems, we
consider a two-dimensional lattice of weakly coupled bond-alternating
chains. CuGeO$_3$ is one example of such a lattice.  The cluster is
shown in Fig. 1, where the axes parallel and perpendicular to the
chain correspond to the $c$- and $b$-axes of CuGeO$_3$, respectively.
The Hamiltonian with an external magnetic field is given by
\begin{eqnarray}
{\cal H}_0=J \sum_{l,j} (1+(-1)^{l+j}\delta)  {\bf S}_{l,j}\cdot{\bf
S}_{l,j+1}
\nonumber\\
          +J_{\perp} \sum_{l,j}{\bf S}_{l,j}\cdot{\bf S}_{l+1,j}
          -{\bf H}\cdot\sum_{l,j}{\bf S}_{l,j},
\label{ham}
\end{eqnarray}
where the index $l$ specifies a chain.  The parameters are fixed as
$J$=1, $\delta =0.1$ and $J_{\perp}$=0.15, realistic for CuGeO$_3$
\cite{Georges}.  We consider the interchain and intrachain DM
interactions (the direction of the vectors ${\bf D}$ is denoted $z$):
\begin{equation}
{\cal H}_{{\rm DM},\perp}= \sum_{l,j}{\bf D}_{\perp l,j}.
\left({\bf S}_{l,j} \times {\bf S}_{l+1,j} \right),
\label{dm1}
\end{equation}
\begin{equation}
{\cal H}_{{\rm DM},\parallel}=  \sum_{l,j}{\bf D}_{\parallel l,j}.
\left({\bf S}_{l,j} \times {\bf S}_{l,j+1} \right).
\label{dm2}
\end{equation}

Using the numerical diagonalization based on the Lanczos algorithm, we
obtain the eigenvalues and eigenvectors of the singlet ground state
and the first excited triplet of the finite cluster described by the
Hamiltonian (\ref{ham}) including the DM term.  Under an external
magnetic field, the triplet splits into three modes with energies
$\Delta _+$, $\Delta _0$ and $\Delta _-$.  To estimate the ESR
intensity for each singlet-triplet transition, we calculate the matrix
elements (\ref{intensity}) for each excitation mode, denoted by
$I_+^{\alpha}$, $I_0^{\alpha}$ and $I_-^{\alpha}$.  The numerical
calculation is performed for the $4\times 4$-spins cluster shown in
Fig. 1 with periodic boundary conditions in both directions.

\begin{figure}
\mbox{\psfig{figure=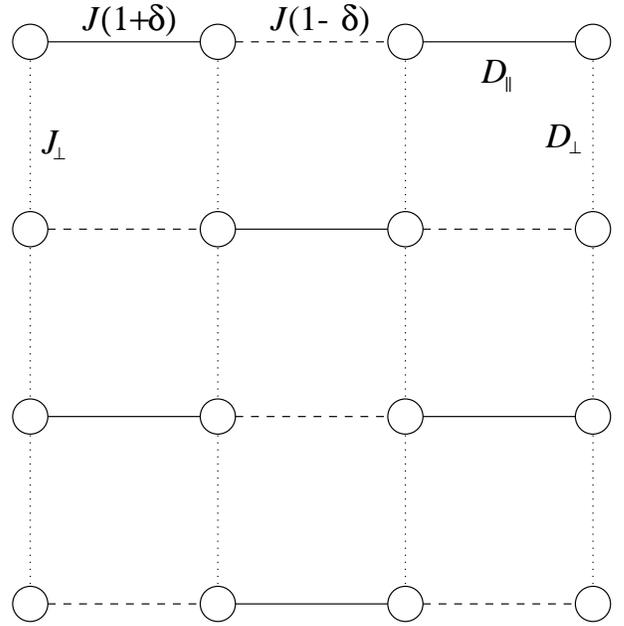,width=80mm,angle=0}}
\vskip 5mm
\caption{
Cluster of 16 spins with DM interaction, used in the exact numerical
diagonalization. The lattice is taken to be similar to that of
CuGeO$_3$.
}
\label{fig1}
\end{figure}
\vskip 3mm

At first, the interchain DM interaction (\ref{dm1}) is investigated.
Since the particular pattern of ${\bf D}$-vectors, $D^z_{\perp
l,j}=(-1)^lD^z_{\perp}$, leads to zero intensity for every
polarization of the wave, we consider the sign-alternating one
$D^z_{\perp l,j}=(-1)^{l+j}D^z_{\perp}$.  In the presence of the DM
interaction with $D^z_{\perp}=0.1$, the ESR intensity is calculated
under various external fields.  To examine the selection rule (i), all
the nonzero intensities are plotted versus the external field $H^z$
which is parallel to ${\bf D}_{\perp}$ in Fig. 2(b).  Figure 2(a) shows
the $H^z$ dependence of the three modes which still have the good
quantum number $S^z_{\rm total}$. Note that the triplet is split
even in zero field because of the DM interaction.  Figure 2(b) shows
that a finite intensity appears only when the polarized field has a
magnetic component perpendicular to the DM vector and the external
field.  Since the magnetization along the external field is
conserved in this case, the intensities are independent of $H^z$,
as shown in Fig. 2(b).  These results agree with rule (i).

\begin{figure}
\mbox{\psfig{figure=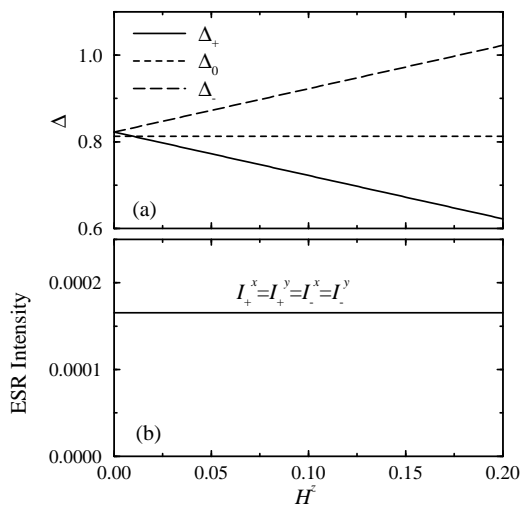,width=80mm,angle=0}}
\vskip 5mm
\caption{
(a) Energy gaps and (b) ESR intensities plotted versus the external
magnetic field parallel to the interchain DM vector.
}
\label{fig2}
\end{figure}
\vskip 3mm

The $D^z$ dependence of the intensity is shown in Fig. 3(a).
The $\ln(I_+^x)$-$\ln(D^z_{\perp})$ plot in Fig. 3(b) is almost linear
with a slope of 2. It indicates the relation

\begin{eqnarray}
I^x=I^y \sim |D^z|^2,
\label{d2}
\end{eqnarray}
which is satisfied in the two-spin system discussed above. (We
omitted the symbol $\perp$ for $D_{\perp}^z$ in eq. (\ref{d2}), because
it
is also held for the intrachain DM interaction.)

\begin{figure}
\mbox{\psfig{figure=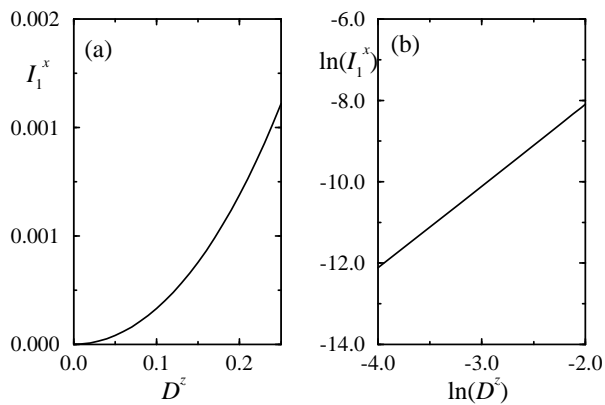,width=80mm,angle=0}}
\vskip 5mm
\caption{
(a) ESR intensity $I_+^x=I_+^y=I_-^x=I_-^y$ versus the amplitude of
the DM vector $D^z$. (b) Plot of $\ln(I_+^x)$-$\ln(D^z)$.
It suggests $I_+^x\sim (D^z)^2$.
}
\label{fig3}
\end{figure}
\vskip 3mm

The results for the external field ($H^x$) perpendicular to the DM
vector are shown in Figs. 4(a) and 4(b).
Figure 4(b) shows that the intensities at the energies $\Delta_+$ or
$\Delta _-$ are finite only if the polarized field ${\bf h}(t)$ has
one component along the external field.
The intensity $I_+^x$ increases as $H^x$ increases.
These behaviors support the validity of
rule (ii) even in many-spin systems.

\begin{figure}
\mbox{\psfig{figure=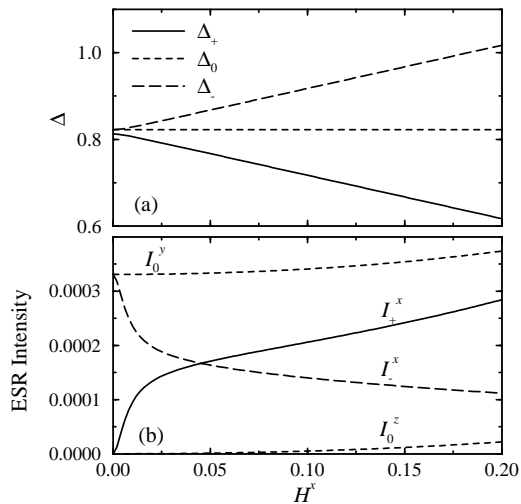,width=80mm,angle=0}}
\vskip 5mm
\caption{
(a) Energy gaps and (b) ESR intensities plotted versus the external
magnetic field perpendicular to the interchain DM vector.
}
\label{fig4}
\end{figure}
\vskip 3mm

Next, the DM coupling along the chain (\ref{dm2}) is investigated.
The intrachain DM interaction gives rise to some nonzero intensities
in the case of a uniform sign, that is,
$D_{\parallel l,j}^z=D_{\parallel}^z$.
The results for such an intrachain DM interaction with
$D_{\parallel}^z=0.1$ are
 presented in Figs. 5 and 6, for the cases of
${\bf H}\parallel {\bf D}$ and ${\bf H}\perp {\bf D}$, respectively.
These results reveal the same behaviors as the interchain DM
interaction for all the wave polarizations,
although the absolute value of the intensity is slightly larger.
Thus, selection rules (i) and (ii) are revealed to be valid also for the
DM coupling along the chain.
These results suggests that the rules do not depend on where
the DM interaction acts.

\begin{figure}
\mbox{\psfig{figure=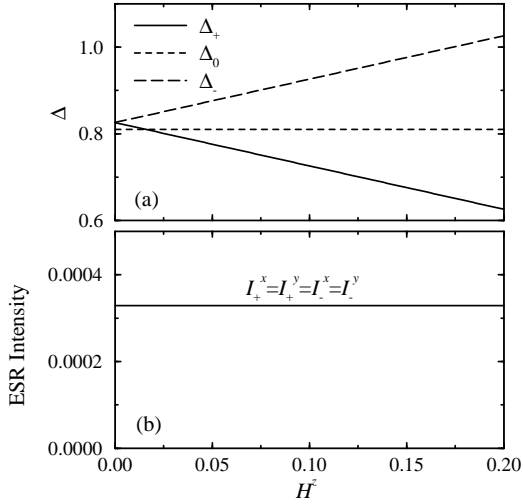,width=80mm,angle=0}}
\vskip 5mm
\caption{
(a) Energy gaps and (b) ESR intensities plotted versus the external
magnetic field parallel to the intrachain DM vector.
}
\label{fig5}
\end{figure}
\vskip 3mm

\begin{figure}
\mbox{\psfig{figure=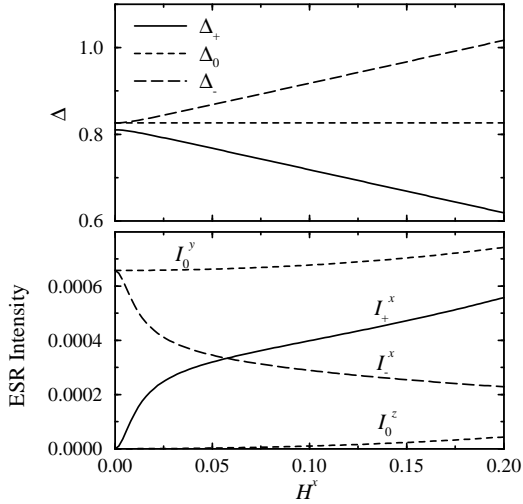,width=80mm,angle=0}}
\vskip 5mm
\caption{
(a) Energy gaps and (b) ESR intensities plotted versus the external
magnetic field perpendicular to the intrachain DM vector.
}
\label{fig6}
\end{figure}
\vskip 3mm

We derived general selection rules (i) and (ii) for a polarized
electromagnetic
wave. In the experiment on CuGeO$_3$, however, the
wave was not polarized, and its wave vector ${\bf k}$ was
oriented along different crystal axes.  In order to compare with these
experimental results, we would average the calculated ESR intensities
over the two components of the magnetic field perpendicular to the wave
vector ${\bf k}$. In the following, we discuss the modified rules for
nonpolarized electromagnetic waves, in the Faraday or Voigt
configurations.

\noindent
(A) Faraday configuration: ${\bf k} \parallel {\bf H}$ \\
Since the magnetic field ${\bf h}(t)$ perpendicular to ${\bf k}$ is
always
perpendicular to the external field, ESR absorption
is allowed only if the rule (i) is satisfied.
The intensity should then be independent of the external field.
It is expected to be largest for ${\bf H} \parallel {\bf D}$ and zero
for ${\bf H} \perp {\bf D}$.  Based on the eq. (\ref{d2}), the
measured intensities for the cases where the external field is
successively along $a$-, $b$- and $c$-axes, would enable us to
determine the direction of the vector ${\bf D}$ in the following way.
The intensity at the energy $\Delta _+$ for ${\bf H}\parallel$$\alpha$
in the Faraday configuration is noted $I^{\alpha}_{\rm (F)}$, so
that the components of ${\bf D}$ should be determined by
\begin{eqnarray}
I^a_{\rm (F)} : I^b_{\rm (F)} : I^c_{\rm (F)}
=(D^a)^2 : (D^b)^2 : (D^c)^2.
\label{ratio}
\end{eqnarray}

\noindent
(B) Voigt configuration: ${\bf k} \perp {\bf H}$ \\ Since ${\bf
h}(t)$, which is perpendicular to ${\bf k}$, has components parallel
and perpendicular to ${\bf H}$, both rules (i) and (ii) can be
realized.  We discuss the two possibilities separately.

\noindent
($\alpha$)${\bf H}\parallel {\bf D}$ (rule (i)):
every intensity is independent of the external field and
the intensities at the energies $\Delta _+$ and $\Delta _-$ are
half that of  the Faraday configuration with ${\bf H}\parallel {\bf
D}$.

\noindent
($\beta$)${\bf H}\perp {\bf D}$ (rule (ii)):
%
%
the ESR intensities at the energies $\Delta _+$ and $\Delta _-$ depend
on the external field.

In a vanishing external magnetic field, symmetry arguments based on the
rotation of the spin operators \cite{Aharony} can be given. In
particular, we explain the vanishing intensity found for a transverse
alternating DM interaction as well as the $D^2$ dependence of the
intensity for the others.

First, consider the alternating transverse DM
interaction $D^z_{\perp l,j}=(-1)^l D_{\perp}^z$. Following Oshikawa
and Affleck
\cite{Oshikawa}, we introduce rotated spin operators $S^{\prime
\pm}_{l,j} = e^{i \pm (-1)^l \theta /2} S^{\pm}_{l,j}$,
%
%
$S^{\prime z}_{l,j} = S^z_{l,j}$ with $\tan \theta =
D^z_{\perp}/J_{\perp}$.
This exactly
maps the Hamiltonian onto an $XXZ$ model
%
%
with alternating coupling
for $S^{\prime}$ operators
for which we can neglect the small second-order anisotropy.
Transforming the Zeeman coupling gives rise to a staggered magnetic
field ${\bf h}^{stagg}(t)$ for rotated operators \cite{Oshikawa}.
The intensity of the transition is given for the rotated variables as

\begin{equation}
I^{x} = \mid \langle g^{\prime} \mid  \cos(\frac{\theta}{2}) \sum_{l,j}
S^{x \prime}_{l,j} +\sin(\frac{\theta}{2}) \sum_{l,j} (-1)^l
 S^{y \prime}_{l,j}  | e^{\prime}
\rangle |^2.
\end{equation}

\noindent
The first term, involving the total spin operator, still
gives a vanishing contribution  because the rotated Hamiltonian is
isotropic. The second term
comes from the field that is staggered in the transverse direction. It
therefore has $q_{\perp}=\pi$ symmetry.
%
%
On the other hand, the lowest-lying excitation and the
ground state have $q_{\perp}=0$ symmetry for an antiferromagnetic
transverse interaction for the lattice of Fig. \ref{fig1}.
\cite{Georges}
This is why the transition
from the ground state to the first gapped excited state is still
forbidden with this DM anisotropy. A transition to the triplet excited
state $q_{\perp}=\pi$ with higher energy is, however, allowed with an
intensity factor
$\sin^2(\theta/2) =
(D^{z}_{\perp}/2J_{\perp})^2$ in the small $D$ limit.\cite{cepas}

Second, consider the case of a constant DM interaction
$D^z_{\parallel l,j}=D^z_{\parallel}$. A rotation to the new set of
operators $S^{\prime \pm}_{l,j} = e^{i \pm \theta j} S^{\pm}_{l,j}$,
$S^{\prime z}_{l,j} = S^z_{l,j}$ (with
%
%
$\tan \theta = D_{\parallel}^z/J$
) also maps the Hamiltonian onto a bond-alternating $XXZ$ model.
In
the rotated frame, the magnetic field ${\bf h}(t)$, however,
transforms into a field which rotates in space with a periodicity $2 \pi
/ \theta$. The intensity is then:

\begin{equation}
I^{x} = \mid \langle g^{\prime} \mid \sum_{l,j} \cos(\theta j) S^{x
\prime}_{l,j}
+ \sin(\theta j) S^{y \prime}_{l,j}  | e^{\prime} \rangle |^2.
\end{equation}

\noindent
Therefore, the intensity of the ESR transition at $q=0$
 induced by uniform DM interaction equals the intensity at small
 $q=\theta$ calculated for a model without DM interaction. For small
$q$, the intensity should behave as $I \sim q^2$
%
%
(the matrix element should be analytic for a
spin-gapped system),
so that $I^{x,y}
 \sim \theta^2=(D_{\parallel}^z/J)^2$, in agreement with the relation
 found numerically.

Finally, we briefly discuss the experimental results of CuGeO$_3$ on
the basis of the above selection rules. We assume that the forbidden
absorption observed is induced by a DM interaction and see what are
the consequences. Considering only the absorption at the energy
$\Delta_+$, the measurement in the Faraday configuration revealed that
$I_{\rm (F)}^c$ is the largest one. Relation (\ref{ratio}) thus
indicates that ${\bf D}$ is along the $c$-axis. The interchain
DM interaction with ${\bf D}_{\perp}$ along the $c$-axis is in fact
the most realistic situation with regards to the crystallographic
structure of CuGeO$_3$. The sign configuration $D^c_{\perp
l,j}=(-1)^{l}D^c_{\perp}$ is, however, expected\cite{cepas,maleyev},
but it leads to zero intensity for every polarization, as proven by
the symmetry argument, and cannot explain the transition observed.
The other patterns for interchain $D^c_{\perp
l,j}=(-1)^{l+j}D^c_{\perp}$ or intrachain $D^c_{\parallel
l,j}=D^c_{\parallel}$ give intensities (the observed intensities would
then indicate an extra small component of ${\bf D}$ along the
$a$-axis), but we cannot conclude from these measurements alone
whether they exist or whether we need other interactions such as
exchange anisotropies or alternation of the $g$-tensor to explain the
ESR absorptions.  In particular, the measured intensities in the
Voigt configuration cannot be compared with the above rules, because
the external field dependence has not been measured yet.  The
observation of increasing intensity with the external field would be
good evidence of the contribution of the DM interaction to the
absorption.
Here, we can make the following observation  for  the case of
NaV$_2$O$_5$:  An absorption at the energy gap
with no strong field dependence has been
observed in
the Faraday configuration \cite{nojiri1} for an external field along the
$a$-axis. If we assume that it is mainly due to DM interaction,
the above rule
(A) would tell us that the DM vector is along the $a$-axis, in agreement
with
the crystallographic structure of V-O-V bonds along the chains.

In summary, we presented two selection rules (i) and (ii) for the ESR
polarized absorption for a spin-gap system in the presence of the DM
interaction. For certain frequencies there may be additional
selection rules, depending on the spatial variations of the
DM  interactions.
We also modified the rules for a nonpolarized electromagnetic
wave for comparison with recent experiments.  It was
found that the intensity depends on the external field in some of
the Voigt configurations.

We thank Professors N. Suzuki and H. Nojiri and
Drs. J. P. Boucher and G. Bouzerar
for fruitful discussions.
The numerical computation was carried out using the
facilities of
the Supercomputer Center, Institute for Solid State Physics,
the University of Tokyo.  This research was supported in part by
Grant-in-Aid for the Scientific Research Fund from the Ministry of
Education, Science, Sports and Culture (No. 11440103).

\widetext
\end{document}